\documentclass[twocolumn,preprintnumbers,amsmath,amssymb]{revtex4}
\usepackage{tabularx,graphicx}

\usepackage{color}
\usepackage{hyperref}
\hypersetup{
    colorlinks=true,
    linkcolor=blue,
    filecolor=blue,      
    urlcolor=blue,
}

\usepackage{color}

\usepackage{ulem}   % to strike things out

\begin{document}
%\documentstyle[aps]{revtex}
%\documentstyle[preprint,aps]{revtex}
%\begin{document}

\newcommand{\beq}{\begin{equation}}
\newcommand{\eeq}{\end{equation}}
\newcommand{\beqn}{\begin{eqnarray}}
\newcommand{\eeqn}{\end{eqnarray}}
\newcommand{\bmath}{\begin{subequations}}
\newcommand{\emath}{\end{subequations}}
\newcommand{\bra}[1]{\langle #1|}
\newcommand{\ket}[1]{|#1\rangle}

%\draft
\title{Comment on arXiv:2312.04495 by M. I. Eremets and coauthors
}

\author{J.  E. Hirsch}
\address{Department of Physics, University of California, San Diego,
La Jolla, CA 92093-0319}
 
 \begin{abstract} 
 In the recently posted arXiv:2312.04495v3 \cite{talv3}, its authors 
 used a computer code written by me  supplied to them by the authors of Ref. \cite{hmtrapped}, and claimed that the results that they obtained  invalidate the results and conclusions
 presented in  the paper J. Supercond. Nov. Mag. 35, 3141-3145 (2022)  by
 F. Marsiglio and myself  \cite{hmtrapped}. Here I point out that the authors' claim
  (i) resulted from improper unauthorized use  of my computer code,
   (ii) is wrong, and (iii) is misleading to the scientific community.
\end{abstract}
 \maketitle

 \section{introduction}
 In Ref.  \cite{hmtrapped}, F. Marsiglio and I analyzed experimental results    on trapped magnetic flux in hydrides under pressure published in
 Ref. \cite{etrappedp} by Minkov,
 Ksenofontov,
 Bud’ko,
Talantsev and
Eremets,
 and we concluded from our analysis that the experimental results indicated that the  residual magnetic moment measured after 
  the external magnetic field was turned off under zero field cooling (ZFC) conditions did not originate in  superconducting currents  \cite{hmtrapped}, hence did not provide
 proof that the materials were superconducting, contradicting the claims of Ref. \cite{etrappedp}.
 
 At a scientific meeting in the Summer of 2023, one of the authors of Ref. \cite{etrappedp} expressed their disagreement with our conclusions in
 Ref.  \cite{hmtrapped}, and their interest to study the computer codes that we  had used for our analysis 
 presented in
 Ref. \cite{hmtrapped}. On September 18, 2023, that author requested by email to F. Marsiglio and myself that we provide the codes used 
 to generate the curves presented in the figures in our paper  Ref.   \cite{hmtrapped}, stating
  {\it ``I am confident that together we will reach a consensus.''}
 
The figures shown in Ref. \cite{hmtrapped} were generated using computer codes written independently
by the two authors of Ref. \cite{hmtrapped}, that were based on the model defined in Ref. 
\cite{hmtrapped},  that yielded identical results. On September 19, 2023, we sent this author one of my computer codes, named $cylinder.f$. We did not include specific instructions on how to use the code, expecting that the author would contact us if  questions arose. At no point did we give  that author nor anybody else permission to publish neither the computer code nor results derived from the computer code.  

%On December 7, 2023, that author together with four coauthors, posted arXiv:2312.04495v1, Ref. \cite{talv1}.
%In it, the authors E.F. Talantsev, V.S. Minkov, V. Ksenofontov, S.L. Bud'ko and M.I. Eremets showed results that they obtained using my computer code,  claiming that the results of my computer code proved our conclusions of Ref. \cite{hmtrapped} wrong.  
%On December 8, 2023, we sent an email to all  the authors of Ref. \cite{talv1}, with subject
%``To authors of arXiv:2312.04495v1'' informing them that they had drawn incorrect conclusions from using my code,
%resulting in statements and figures in their paper that are wrong and misleading to the scientific community. 
%We also pointed out to the authors that we had not authorized them to publish results obtained using my code.
%None of the authors responded. In the ensuing days, they posted new versions of their paper \cite{talv2,talv3}
%that did not correct the problem.
%
%Our email of December 8 to the authors   was copied to arXiv moderation. In addition,  
%we wrote to arXiv moderation and to arXiv's scientific Director Steinn Sigurdsson also on December 8, informing that
%Ref. \cite{talv1} had   posted results obtained from my code that were incorrect and
%profoundly misleading to readers, and that at no point did we grant permission for  publication by others of results obtained
%using  my code, and requesting that 
%arXiv withdraws arXiv:2312.04495v1.
%
%Arxiv responded on December 11, 2023 that it would take no action, and  has  taken
%no action as of today, December 18, 2023. That is the reason for this Comment.

       \begin{figure*} [t]
 \resizebox{18.5cm}{!}{\includegraphics[width=6cm]{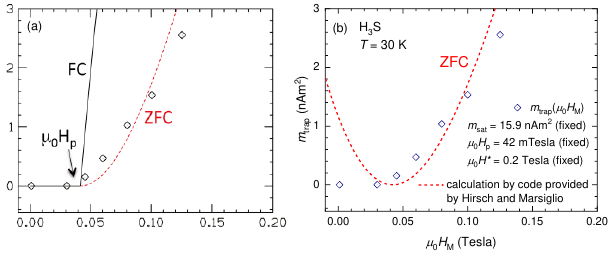}} 
 \caption {Panels (a) and (b) of Fig. 2 of Refs. \cite{talv1,talv2,talv3}. 
The figure caption of Refs. \cite{talv1,talv2,talv3}, as it refers to these panels, reads 
{\it ``Experimental ZFC $m(B_{appl}, T=30K)$ dataset measured in highly compressed $H_3S$, together with reproduced simulations and standard data fitting. (a) The original Figure 5 from Ref.1; (b) the reproduced full simulation curve utilizing the original simulation code, the reported fixed 
$\mu_0 H_p, \mu_0 H^*, m_{sat}$, and $\mu_0 H_{max}$    parameters and the aforementioned authors´ manipulations demonstrating the unphysical anomaly within the Meissner state...Open rhombuses represent experimental data, black and red curves represent simulation and fitting curves.''} Here, Ref. 1 is Ref. \cite{etrappedp} and ``the aforementioned authors'' are the authors of Ref. \cite{hmtrapped}.
}
 \label{figure1}
 \end{figure*} 
 
 \section{the physics and the computer code}
 In Ref. \cite{hmtrapped}, we modeled what the trapped magnetic field in a hard superconductor would be when a magnetic field
 $H_M$ is applied at low temperatures and subsequently removed, i.e. zero field cooling  (ZFC) protocol. We used the well-known 
 Bean model, that assumes
 a critical current independent of magnetic field. 
 
 The model has two important parameters: (1) The threshold magnetic field $H_p$, below which the applied magnetic field does not penetrate
 and hence cannot be trapped, and the field $H^*$: for applied magnetic field $H_M=H^*$,   the applied field reaches the center of the sample. $H^*$  is proportional to
 the critical current density. The threshold magnetic field $H_p$ is the lower critical field $H_{c1}$ if the sample is a long cylinder,
 and for other sample shapes it is the lower critical field $H_{c1}$ corrected for demagnetization. In any case, 
 for applied magnetic field $H_M<H_p$ the applied magnetic field does not penetrate the sample, hence cannot be trapped,
 hence when the external magnetic field is removed the resulting  magnetic moment is exactly zero.
 
 All the figures shown in our paper Ref. \cite{hmtrapped} that show results of our model, namely Fig. 3, Fig 4 and Fig. 5, and their insets,
 show that the trapped magnetic moment predicted by our model for $H_M<H_p$ is exactly $zero$.
 
 In their paper arXiv:2312.04495v1,v2,v3 \cite{talv1,talv2,talv3} the authors show in their Fig. 2 the two panels shown in Fig. 1 here.
 The left panel of Fig. 1 shows the inset of Fig. 5 of our paper Ref. \cite{hmtrapped}, the red dashed curve on the left panel of Fig. 1 shows the prediction
 of our model for zero field cooling (ZFC), as published in our paper Ref. \cite{hmtrapped}. The red curve reaches zero at magnetic field $H_M=H_p=0.042T$ and does not
 continue for $H_M<H_p$ because in that range there cannot be a trapped moment under ZFC since the applied field never penetrated
 the sample, as explained above.
 
 The right panel of Fig. 1 show what the authors of Refs. \cite{talv1,talv2,talv3} obtained when they ran the computer code
 $cylinder.f$ provided by us to them and plotted the numbers that the computer code printed in unit 12. Those numbers joined
 by a smooth line result in the dashed red line on the right  panel of Fig. 1.
 
 It can be seen that the dashed red curves are identical on the left and right panels of Fig. 1
 in the field range $H_M \ge H_p$, with $H_p=0.042T$.  Instead, for $H_M<H_p$ there is no
 dashed red curve on the left panel, while on the right panel there is a dashed red curve that rises
 as $H_M$ becomes lower.
 
 What is the reason for this discrepancy? On the left panel there is no dashed red curve for $H_M<H_p$ because there can
 be no trapped magnetic moment if the magnetic field didn't penetrate, as explained above. 
 On the right panel there is a dashed red curve for $H_M<H_p$ because the authors of Refs. \cite{talv1,talv2,talv3}
 took the numbers from the output of the computer code $cylinder.f$ in the region 
 $H_p<H_M$ and plotted them and connected them with a dashed red line.
 
 However, those numbers don't have any physical meaning and should not be interpreted as
 having any physical meaning. Anybody that understands the computer code $cylinder.f$ and how to use it, and understands
 the physics of the model used in our Ref. \cite{hmtrapped} to describe trapped flux in hard superconductors, knows this.
 The authors of Refs.  \cite{talv1,talv2,talv3}, because they ran a computer code that did not belong to them,
 and used it without checking with the authors of the code how the code should be used and how the numbers printed out by the code should be interpreted, apparently  don't know this.
 
 From the upturn in the dashed red curve seen on the right panel of Fig. 1 for field $H_M<H_p$, and similar upturns shown
 in Figs. 1(a), (1b) of Refs.  \cite{talv1,talv2,talv3} also for $H_M<H_p$, the authors of Refs.  \cite{talv1,talv2,talv3}
 concluded that:

\begin{itemize}

\item
 {\it ``Significantly, the Hirsch-Marsiglio model does not describe the Meissner regime''},
 \item
{\it ``the model predicts a significant positive magnetic moment for ZFC''}, 
\item
{\it ``This feature of the proposed model was not discussed by the authors and obviously contradicts the physics for the trapped magnetic flux in superconductors''}, 
\item
{\it ``they simulated the $m(B_{appl})$   curves by implementing the following step-by-step procedure...
manually hide/delete all simulated points within the Meissner state, i.e. for $B_{appl} \le \mu_0 H_p$}

\item{\it ``we conclude that the authors in their Figures 3-5 hid/deleted (without reporting this) parts of their simulated datasets that disagree with the Meissner state. This potentially allowed them to conceal the issue that their model and computer code do not adequately describe the Meissner state.''}

\item
{\it ``The simulations conducted by the
authors involve ...  and unjustifiably deleting parts of the simulation dataset.''}

\item Caption of Fig. 2:
{\it ``...the aforementioned authors´ manipulations demonstrating the unphysical anomaly within the Meissner state''}

 \item Abstract: {\it ``Hirsch and Marsiglio, in their recent publication (J. Supercond. Nov. Mag. 35, 3141–3145, 2022), assert that experimental data on the trapping of magnetic flux by hydrogen-rich compounds clearly demonstrate the absence of superconductivity in hydrides at high pressures. We argue that this assertion is incorrect, as it relies on the wrong model coupled with selective manipulations (hide/delete) of calculated datasets...''}.
\end{itemize}

These conclusions of Refs.   \cite{talv1,talv2,talv3}  are (i) wrong, (ii) misleading to the scientific community, and 
(iii) were obtained  using a computer
code that did not belong to the authors  of Refs.   \cite{talv1,talv2,talv3},   in a way that was incorrect and  inconsistent with the way the computer code
was designed to be used,  without authorization from the code's authors, and without checking with the code authors whether
the code was used properly.
 
 \section{summary}

 In summary, Refs.  \cite{talv1,talv2,talv3} used   in an incorrect way a computer code written by me supplied by 
 F. Marsiglio and me  to one of the authors of Ref. \cite{etrappedp} responding to their request
 with the stated goal   {\it ``I am confident that together we will reach a consensus.''},
drew an incorrect conclusion from it, and posted Refs.   \cite{talv1,talv2,talv3}, that contain several explicit statements misrepresenting our work
Ref. \cite{hmtrapped}, as discussed in this Comment. Namely, the incorrect conclusion that the model used in
 Ref. \cite{hmtrapped} yields  the 
 unphysical conclusion that a trapped magnetic moment appears when a magnetic field smaller than
 the lower critical field is applied to the sample and then removed, in direct contradiction with the analysis, discussion and results
 presented in our paper Ref. \cite{hmtrapped}. The authors of Refs.  \cite{talv1,talv2,talv3}
 did not request nor receive permission from us to publish results derived from the code that we sent them, nor did they send us their paper in advance of posting it (nor thereafter) to request our comments. 
% When we
% notified them more than 10 days ago that they had used our computer code incorrectly and asked that they
% replace the paper  they had posted correcting the error, they did not respond, ignored our request, and posted two subsequent
% versions of their paper  \cite{talv2,talv3} with essentially the same content. 
%This is in violation of
% normal scientific practice.
 
 In a separate paper, F. Marsiglio and I will address other issues about our paper  Ref. \cite{hmtrapped}
 that  were raised in arXiv:2312.04495.

 Note: the original version of this Comment, submitted to arXiv on 12/19/2023,   can be found in Ref. \cite{original}.


\begin{references}
      

      
      \bibitem{talv3} E.F. Talantsev, V.S. Minkov, V. Ksenofontov, S.L. Bud'ko and M.I. Eremets,
      ``Is MgB2 a superconductor?
Comment on “Evidence Against Superconductivity in Flux Trapping Experiments on Hydrides Under High Pressure” [J. E. Hirsch and F. Marsiglio in J. Supercond. Nov. Mag. 35, 3141–3145 (2022)]'',
      \href{https://arxiv.org/abs/2312.04495}{arXiv:2312.04495v3, December 13, 2023}.
      
      \bibitem{hmtrapped} J. E. Hirsch and F. Marsiglio,
``Evidence Against Superconductivity in Flux Trapping Experiments on Hydrides Under High Pressure'', \href{https://link.springer.com/article/10.1007/s10948-022-06365-8}
{J Supercond Nov Magn 2022; 35: 3141–3145}.



\bibitem{etrappedp}     V. S. Minkov,
V. Ksenofontov,
S. L. Bud’ko,
E. F. Talantsev and
M. I. Eremets,
``Magnetic flux trapping in hydrogen-rich high-temperature superconductors'',
\href{https://www.nature.com/articles/s41567-023-02089-1}{Nat. Phys. (2023)}.

   \bibitem{talv1} E.F. Talantsev, V.S. Minkov, V. Ksenofontov, S.L. Bud'ko and M.I. Eremets,
      ``Is MgB2 a superconductor?
Comment on “Evidence Against Superconductivity in Flux Trapping Experiments on Hydrides Under High Pressure” [J. E. Hirsch and F. Marsiglio in J. Supercond. Nov. Mag. 35, 3141–3145 (2022)]'',
      \href{arXiv:2312.04495v1}{arXiv:2312.04495v1, December 7, 2023}.
      
         \bibitem{talv2} E.F. Talantsev, V.S. Minkov, V. Ksenofontov, S.L. Bud'ko and M.I. Eremets,
      ``Is MgB2 a superconductor?
Comment on “Evidence Against Superconductivity in Flux Trapping Experiments on Hydrides Under High Pressure” [J. E. Hirsch and F. Marsiglio in J. Supercond. Nov. Mag. 35, 3141–3145 (2022)]'',
      \href{arXiv:2312.04495v2}{arXiv:2312.04495v2, December 10, 2023}.
      
            \bibitem{original} J. E. Hirsch, 
      ``Comment on arXiv:2312.04495 by M. I. Eremets and coauthors'',
      \href{https://osf.io/preprints/osf/p29ht}{https://doi.org/10.31219/osf.io/p29ht}  (2023).

%\bibitem{beanrmp} C. P. Bean, ``Magnetization of High-Field Superconductors'',
%\href{https://journals.aps.org/rmp/abstract/10.1103/RevModPhys.36.31}{Rev. Mod. Phys. 36, 31  (1964)}.

%          \bibitem{e2015} A.P. Drozdov, M.I. Eremets, I. A.Troyan, V. Ksenofontov  and S. I. Shylin,
%     `Conventional superconductivity at 203 kelvin at high pressures in the sulfur hydride system',
%     \href{https://www.nature.com/articles/nature14964}{Nature 525, 73-76 (2015)}.
%     
%                 \bibitem{troyan} I. A. Troyan et al,
%     ``High-temperature superconductivity in hydrides'',
%     \href{https://ufn.ru/en/articles/2022/7/h/}{Phys. Usp. 65, 748  (2022)}.
%     
%
     
%     \bibitem{nonstandard}
%J. E. Hirsch and F. Marsiglio, ``Nonstandard superconductivity or no superconductivity in hydrides under high pressure'', \href{https://journals.aps.org/prb/abstract/10.1103/PhysRevB.103.134505}{Phys. Rev. B 103, 134505 (2021)}.
%
% \bibitem{hmmre} J. E. Hirsch and F. Marsiglio, ``Clear evidence against superconductivity in hydrides under high pressure'',
%\href{https://aip.scitation.org/doi/10.1063/5.0091404}{Matter and Radiation at Extremes 7, 058401 (2022)}.
%     
%          \bibitem{pers}
%     J. E. Hirsch, ``Are hydrides under high pressure high temperature superconductors?'',
%     \href{https://doi.org/10.1093/nsr/nwad174}{National Science Review, nwad174 (2023)}
%and references therein.

%         \bibitem{retraction1}       E. Snider et al,
%     ``Retraction Note: Room-temperature superconductivity in a
%carbonaceous sulfur hydride'',
%     \href{https://doi.org/10.1038/s41586-022-05294-9}{Nature  610, 804 (2022)}.
%         
%             \bibitem{retraction2} N. Dasenbrock-Gammon et al,
%             ``Retraction Note: Evidence of near-ambient superconductivity in a N-doped lutetium hydride'',
%             \href{https://www.nature.com/articles/s41586-023-06774-2}
%             {Nature 624, 460 (2023). https://doi.org/10.1038/s41586-023-06774-2}.
%     
%     \bibitem{retraction3}
%     E. Snider et al,
%     ``Expression of Concern: Synthesis of Yttrium Superhydride Superconductor with a Transition Temperature up to 262 K by Catalytic Hydrogenation at High Pressures [Phys. Rev. Lett. 126, 117003 (2021)]'',
%     \href{https://journals.aps.org/prl/abstract/10.1103/PhysRevLett.131.239902}{Phys. Rev. Lett. 131, 239902  (2023)}.
%     
%       \bibitem{e2021p} V. S. Minkov, S. L. Bud’ko, F. F. Balakirev, V. B. Prakapenka, S. Chariton, R. J. Husband, H. P. Liermann and M. I. Eremets, ``Magnetic field screening in hydrogen-rich high-temperature superconductors'',
%\href{https://www.nature.com/articles/s41467-022-30782-x} {Nat Commun 13, 3194 (2022)}.
%
%
%
%\bibitem{hmscreening}  J. E. Hirsch and F. Marsiglio, ``On Magnetic Field Screening and Expulsion in Hydride Superconductors'',
%\href{https://link.springer.com/article/10.1007/s10948-023-06569-6}{J Supercond Nov Magn 2023;  36: 1257–1261}.
%
%\bibitem{unpulling} J. E. Hirsch and F. Marsiglio, 
%``On magnetic field screening and trapping in hydrogen-rich high-temperature superconductors: unpulling the wool over readers' eyes'',
%\href{https://arxiv.org/abs/2309.02683}{arXiv:2309.02683 (2023)}, \href{https://link.springer.com/article/10.1007/s10948-023-06622-4}
%        { J Supercond Nov Magn (2023) https://doi.org/10.1007/s10948-023-06622-4}.
%        
%        
%\bibitem{bending} J. E. Hirsch, 
%``Can linear transformations bend a straight line? Comment on Author Correction to Magnetic field screening in hydrogen-rich high-temperature superconductors'', \href{https://www.sciencedirect.com/science/article/pii/S0921453423001910}{Physica C 616, 1354400) (2024)}.
% 
%
%        
%
%\bibitem{correction} V. S. Minkov, S. L. Bud’ko, F. F. Balakirev, V. B. Prakapenka, S. Chariton, R. J. Husband, H. P. Liermann and M. I. Eremets,  ``Author Correction: Magnetic field screening in hydrogen-rich high-temperature superconductors'',
%\href{https://www.nature.com/articles/s41467-023-40837-2}{Nat Commun 14, 5322 (2023)}.
% 
%
%\bibitem{hysteresis} J. E. Hirsch, 
%"Hysteresis loops in measurements of the magnetic moment of hydrides under high pressure: Implications for superconductivity'',
%\href{https://www.sciencedirect.com/science/article/pii/S0921453424000145}{Physica C journal pre-proofs  1354449},
%14 January 2024.
%``
%
%

                 \end{references}
 \end{document}